\documentclass[fleqn,10pt]{wlscirep}
\usepackage[utf8]{inputenc}
\usepackage[T1]{fontenc}
\usepackage{bm}
\usepackage{siunitx}
\usepackage{hyperref}
\usepackage{subfiles}
\usepackage{caption}
\usepackage{authblk}
\usepackage{glossaries}

\title{Parameter Efficient Machine Unlearning on Hybrid Resistive Memory based Compute-in-Memory Accelerators}

\author[1,3,10]{Ning Lin}
\author[1,2,5,10]{Jichang Yang}
\author[1,2,5,10]{Yangu He}
\author[1]{Zijian Ye}
\author[1]{Kwun Hang Wong}
\author[1,5]{Xinyuan Zhang}
\author[1,5]{Songqi Wang}
\author[2]{Zihao Li}
\author[3]{Yuxi Chen}
\author[3]{Jiajia Zha}
\author[6]{Wenxing Li}
\author[1,5]{Yi Li}
\author[8,9]{Kemi Xu}
\author[7]{Leo Yu Zhang}
\author[6]{Xiaoming Chen}
\author[4]{Dashan Shang}
\author[3]{Chaoliang Tan}
\author[1,5,*]{Han Wang}
\author[1,*]{Xiaojuan Qi} 
\author[2,*]{Zhongrui Wang}

\affil[1]{Department of Electrical and Computer Engineering, the University of Hong Kong, Hong Kong, China}
\affil[2]{The School of Microelectronics, Southern University of Science and Technology, Shenzhen 518055, China}
\affil[3]{Department of Electrical Engineering, City University of Hong Kong, Hong Kong,999077, China}
\affil[4]{Institute of Microelectronics, Chinese Academy of Sciences, Beijing 100029, China}
\affil[5]{Centre for Advanced Semiconductors and Integrated Circuits, the University of Hong Kong, Hong Kong, China}
\affil[6]{Institute of Computing Technology, Chinese Academy of Sciences, Beijing 100190, China}
\affil[7]{The School of Information and Communication Technology, Griffith University, QLD 4215, Australia}
\affil[8]{MIIT Key Laboratory of Complex-field Intelligent Sensing, School of Optics and Photonics, Beijing Institute of Technology, Beijing 100081, China}
\affil[9]{Center for Photonic Quantum Precision Measurement, Advanced Research Institute of Multidisciplinary Science, Beijing Institute of Technology, Beijing 100081, China}
\affil[10]{These authors contributed equally.}
\affil[*]{e-mail: hanwang6@hku.hk; xjqi@eee.hku.hk; zrwang@eee.hku.hk}

\begin{abstract}
Resistive memory compute-in-memory accelerators provide energy efficient analogue matrix vector multiplication for neural network inference, but frequent reprogramming of analogue weights remains costly because of device variability and iterative write and verify operations. This limitation hinders their use in edge model adaptation, including approximate machine unlearning and continual learning, where model parameters may need to be updated repeatedly in response to data deletion requests or newly arriving tasks. Here we present a co-design approach across hardware and software that maps frozen pretrained weights to analogue resistive memory arrays while placing trainable low rank adaptation branches in SRAM connected digital compute. By using LoRA style parameter efficient updates, the proposed scheme confines adaptation to a small set of digital parameters and avoids repeated reprogramming of the analogue backbone.
To our knowledge, this work provides the first experimental demonstration of approximate machine unlearning on a fabricated resistive memory CIM accelerator. We validate the framework on a 180 nm 128$\times$128 1T1R resistive-memory macro for face recognition, and through circuit-accurate simulations for speaker authentication and stylized image generation tasks, owing to the substantial model sizes involved. Compared with a baseline that directly updates analog weights, our hybrid mapping reduces analog training/update cost by up to 148$\times$, on-chip deployment overhead by up to 388$\times$, and inference energy by up to 59$\times$, while preserving competitive task performance.
These results show that hybrid analogue-digital LoRA mapping can enable efficient post-deployment adaptation on RM-CIM hardware, although formal machine-unlearning guarantees and large-scale system integration remain open challenges.
\end{abstract}

\begin{document}

\flushbottom
\maketitle

\thispagestyle{empty}

\section*{Introduction}

\begin{figure}[!t]
\centering
\includegraphics[width=1.0\linewidth]{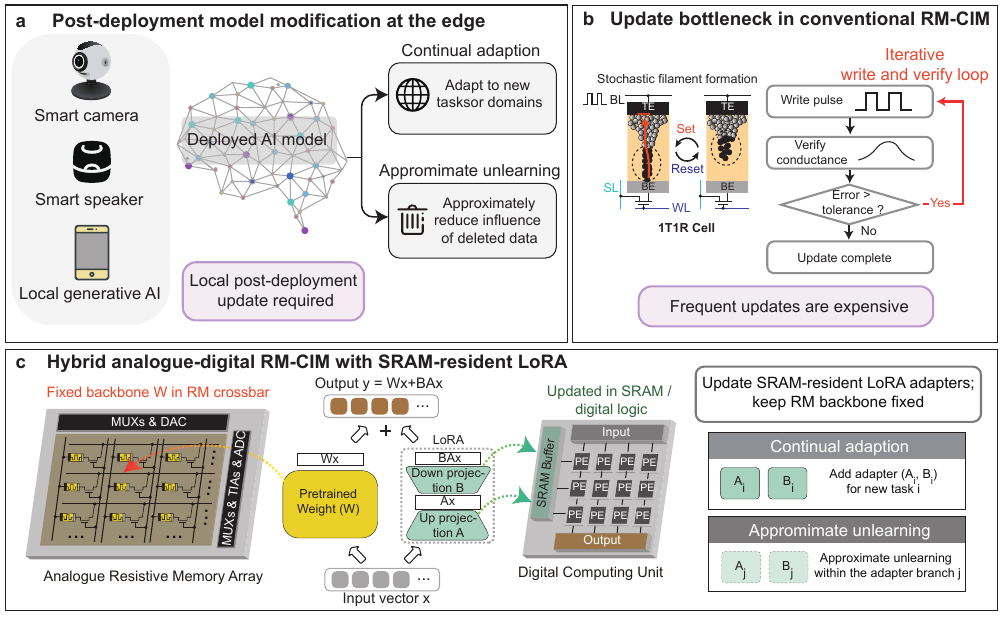}
\caption{\textbf{Hardware--software co-design for LoRA-based approximate machine unlearning and continual learning on a hybrid analogue--digital RM-CIM system.} 
\textbf{a}, Edge AI models deployed on personal devices may require local post-deployment modification to support continual adaptation to new tasks or domains and approximate unlearning of deleted or sensitive data.
\textbf{b}, In conventional RM-CIM accelerators, stochastic and device-dependent resistive-memory programming requires iterative write-and-verify operations to set mapped analogue weights to target conductance states. Frequent weight updates therefore incur substantial energy and latency overheads and may accumulate programming errors, creating an update bottleneck for adaptive learning.
\textbf{c}, Proposed hybrid analogue--digital RM-CIM architecture with SRAM-resident LoRA adapters. The fixed pretrained backbone weights are mapped to analogue resistive-memory crossbars for efficient matrix--vector multiplication, while lightweight LoRA branches are stored and updated in SRAM-connected digital compute units. Post-deployment continual adaptation and approximate unlearning are therefore confined to the digital LoRA parameters, avoiding repeated reprogramming of the analogue RM backbone.
}
\label{fig1}
\end{figure}

Edge AI systems are increasingly deployed in personal devices such as cameras, speakers and mobile generators, where models may need to change after deployment as users, tasks or privacy requirements evolve.
Personal and embedded devices may need to incorporate user-specific data, adapt to newly arriving tasks, correct undesirable behaviours or reduce the influence of data associated with deletion requests. 
Cloud-based retraining followed by redeployment is often impractical when data are sensitive, communication is costly or rapid local adaptation is required. 
Efficient on-device model modification is therefore becoming an important requirement for continual learning and approximate machine unlearning, as illustrated in Fig.~\ref{fig1}a.

Resistive-memory compute-in-memory (RM-CIM) accelerators are promising for energy-efficient edge inference. 
By storing neural-network weights as analogue conductance states and performing matrix--vector multiplication directly inside dense crossbar arrays, they reduce the data movement that dominates conventional von Neumann architectures~\cite{lanza2025growing,wan2022compute,yao2020fully,liu2022optoelectronic}. 
However, this advantage relies on the mapped weights remaining largely fixed. 
Programming resistive-memory devices is stochastic and device-dependent, and typically requires iterative write-and-verify operations to reach a target conductance level. 
Frequent weight updates therefore incur substantial energy and latency overheads and may accumulate programming errors~\cite{rao2023thousands,song2024programming,zuo2025precise,yangdong2025ultrahigh}. 
This creates a practical mismatch between adaptive learning algorithms, which require repeated model modification, and RM-CIM hardware, which favours stable analogue weights, as illustrated in Fig.~\ref{fig1}b.

This mismatch motivates a hardware--algorithm partition in which the large analogue backbone remains fixed while post-deployment changes are confined to a compact and easily updateable parameter space, as illustrated in Fig.~\ref{fig1}c. 
Prior RM-CIM studies have mainly addressed inference efficiency, mapping accuracy, device variation, programming precision and, more recently, hardware security of stored models~\cite{wang2024safe,liu2026privacy,lin2025guarder,ahmed2026mof,wu2026hardware,yue2025physical}. Less attention has been paid to how a deployed RM-CIM model can be repeatedly modified after deployment without incurring the cost of reprogramming analogue weights.
Here, we address the complementary problem of how a deployed RM-CIM model can be efficiently personalized, continually adapted or approximately unlearned without repeatedly reprogramming the analogue weight arrays. 
The objective is not only to accelerate inference, but also to make model modification compatible with the physical constraints of RM-CIM hardware.

Parameter-efficient adaptation provides a natural mechanism to realize this partition. 
Low-rank adaptation (LoRA) freezes a pretrained weight matrix and learns only a compact low-rank update~\cite{hu2022lora}. 
In software, adapter-based methods have enabled efficient fine-tuning and continual learning with far fewer trainable parameters than full-model updating~\cite{wistuba2023continual}. 
For RM-CIM systems, this suggests a hybrid analogue--digital mapping: the large and relatively stable pretrained backbone is deployed in analogue resistive-memory arrays, whereas the small adaptive branch is stored in SRAM and updated by digital logic. 
Keeping the adapter separate from the analogue backbone avoids repeated conductance reprogramming and allows task- or user-specific behaviours to be introduced, switched, updated or approximately removed with low overhead.

Here we propose and experimentally evaluate such a hybrid analogue--digital RM-CIM mapping for parameter-efficient continual adaptation and approximate machine unlearning. 
Frozen pretrained weights are mapped to analogue RM crossbars, while trainable LoRA branches are implemented in SRAM-connected digital compute. 
This co-design confines post-deployment updates to a small number of digital parameters, thereby reducing analogue write cost while preserving the energy-efficiency benefit of RM-CIM inference. 
The unlearning considered in this work is approximate and targets behaviours represented in the adaptation branch, rather than certified removal of arbitrary information from the entire pretrained model.

We validate the approach experimentally on a fabricated 180 nm 128$\times$128 1T1R RM-CIM macro for the face identity recognition workload, and through circuit-accurate simulations for speaker authentication and stylized image generation.
Compared with a baseline that reprograms all analogue parameters, updating only the SRAM-resident LoRA parameters reduces training cost by approximately 27.51$\times$--147.76$\times$, reduces on-chip deployment overhead by approximately 69.68$\times$--387.95$\times$, and reduces inference energy by approximately 13.07$\times$--59.44$\times$, while maintaining competitive task performance. 
These results demonstrate hybrid analogue--digital LoRA mapping as a practical route towards hardware-efficient model modification on fabricated RM-CIM accelerators.

\begin{figure}[!t]
\centering
\includegraphics[width=1.0\linewidth]{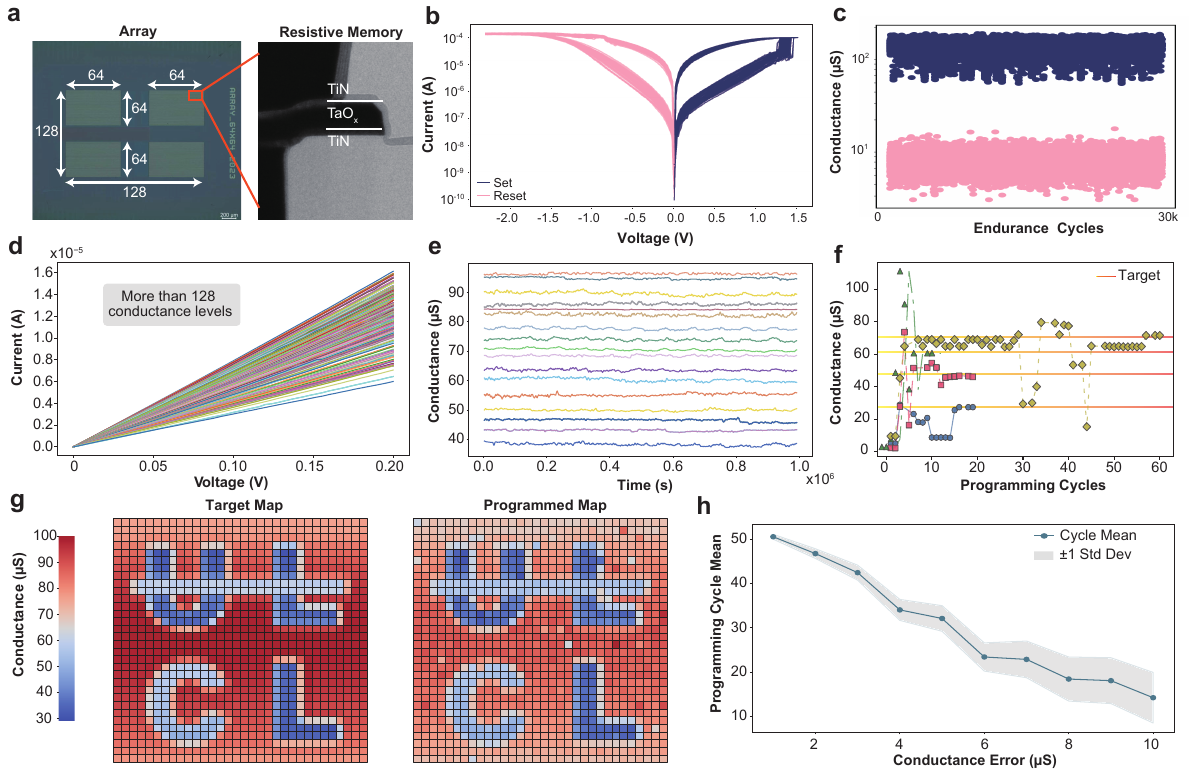}
\caption{\textbf{RM device characteristics for programming.}
\textbf{a,} Optical micrograph of the 128 $\times$ 128 RM array and the RM stack.
\textbf{b,} Quasi-static $I$-$V$ sweeps of an RM cell over 50 cycles, showing repeatable bipolar resistive switching.
\textbf{c,} Endurance of an RM cell over 30{,}000 SET/RESET cycles.
\textbf{d,} Single-shot SET programming of an RM cell into more than 128 conductance levels by varying the programming voltage.
\textbf{e,} Time evolution of the programmed conductance over $10^{6}$ s, demonstrating stable readout.
\textbf{f,} Programming trajectories for four RM cells with conductances tuned towards different target values.
\textbf{g,} Target and programmed conductance maps encoding `UL' (machine unlearning) and `CL' (continual learning), programmed using a halting criterion that stops the write operation when the programming error is within a 2~$\mu$S tolerance.
\textbf{h,} Relationship between the number of programming cycles per cell and the conductance error after programming (solid line, mean; shaded area, standard deviation).
}
\label{fig2}
\end{figure}

\begin{figure}[!t]
\centering
\includegraphics[width=1.0\linewidth]{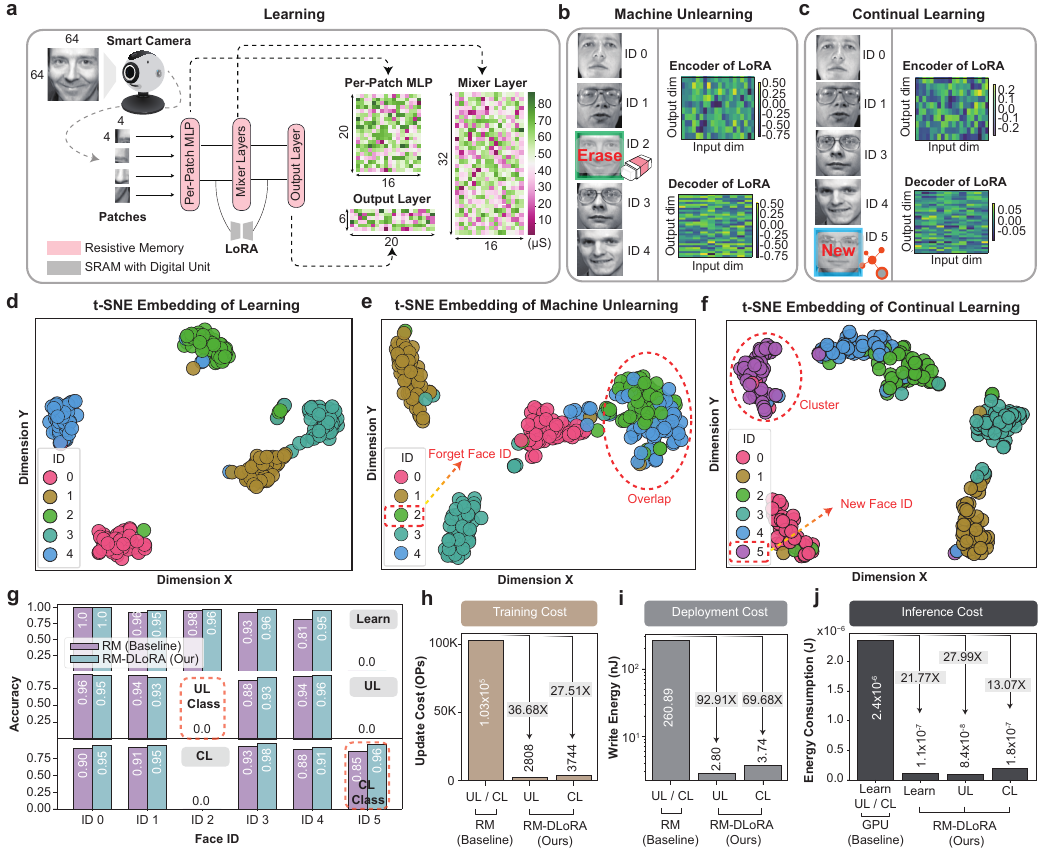}
\caption{\textbf{Learning, machine unlearning (UL) and continual learning (CL) for face classification on the Olivetti faces dataset.} \textbf{a,} Schematic of the MLP-Mixer architecture mapped onto RM array for initial learning. 
\textbf{b,} Machine unlearning of face ID~2 using exemplar LoRA encoder and decoder modules. 
\textbf{c,} Continual learning of face ID~5 using exemplar LoRA encoder and decoder modules. 
\textbf{d-f,} Two-dimensional t-distributed stochastic neighbour embedding (t-SNE) visualizations of the embedding space after learning (\textbf{d}), machine unlearning (\textbf{e}) and continual learning (\textbf{f}). 
\textbf{g,} Classification accuracy obtained with the our hybrid RM with digital LoRA architecture (RM-DLoRA) and with fully analogue RM reprogramming baseline method, showing higher accuracy for RM-DLoRA owing to the avoidance of repeated analogue reprogramming in learn, machine unlearning (UL) and continual learning (CL) tasks. 
\textbf{h,} Training update cost for full-parameter fine-tuning on RM and RM-DLoRA, our method reducing training cost by about 36.68$\times$ and 27.51$\times$ in UL and CL tasks. 
\textbf{i,} Write-energy consumption for full-parameter fine-tuning on RM and RM-DLoRA, our method reducing write energy by about 92.91$\times$ and 69.68$\times$ in UL and CL tasks.
\textbf{j,} Inference energy for RM-DLoRA and a GPU baseline, with RM-DLoRA achieving about 21.77$\times$, 27.99$\times$ and 13.07$\times$ lower energy consumption in learning (Learn), UL and CL tasks.}
\label{fig3}
\end{figure}

\section*{Results}

\subsection*{RM Device Characterization and Conductance Programming}

Fig.~\ref{fig2}a presents optical micrographs of the $128 \times 128$ RM array and TEM images of the TiN/TaO$_x$/TiN RM stack (see Supplementary Fig.~1 for the detailed chip architecture).
Panels (b)-(f) of Fig.~\ref{fig2} summarise the electrical characteristics and multilevel read/write capabilities of the RM devices. Quasi-static $I$--$V$ sweeps over 50 cycles (Fig.~\ref{fig2}b) exhibit highly uniform bipolar resistive switching. Robust endurance is observed over 30{,}000 cycles of SET and RESET operations (Fig.~\ref{fig2}c). 
Single-shot SET programming yields more than 128 approximately linearly spaced conductance levels (Fig.~\ref{fig2}d), demonstrating the strong multilevel capability of the RM cells.
Repeated readout of 17 RM cells with distinct conductance levels (Fig.~\ref{fig2}e) demonstrates stable programmed conductances over $10^6$~s, with minimal temporal fluctuations and read noise, indicating excellent read stability.
Fig.~\ref{fig2}f shows programming trajectories for four RM cells targeting distinct conductance values, each requiring tens of operations. 

Panels (g)-(h) of Fig.~\ref{fig2} highlight the challenges inherent in RM programming. 
In Fig.~\ref{fig2}g, target conductance maps encoding the `CL' (continual learning) and `UL' (machine unlearning) patterns are programmed using a halting criterion that stops the write operation when the programming error is within a 2~$\mu$S tolerance, resulting in only a small residual conductance error and programmed maps that closely match the targets.

Achieving this precision, however, requires iterative tuning. Fig.~\ref{fig2}h quantifies this trade-off by plotting the mean number of programming cycles per cell against the conductance error (shaded area, standard deviation); higher precision demands substantially more iterations. For example, achieving an error below 1~$\mu$S requires approximately 50 cycles per cell on average.

\subsection*{Face Recognition}

Face recognition in smart cameras is highly privacy-sensitive, motivating mechanisms for controlled identity removal and efficient model adaptation. After deployment, such systems must accommodate personnel turnover by adding or removing identities from the recognition database. Accordingly, the underlying model should support approximate machine unlearning of removed identities and continual learning of newly added ones, while preserving recognition performance. We demonstrate this capability on face-recognition tasks using an RM analogue--digital system equipped with rank $r=6$ LoRA adapters, evaluated on the Olivetti faces dataset~\cite{samaria1994parameterisation}.

Fig.~\ref{fig3}a-c illustrate the model architecture and parameter transformations during learning, machine unlearning and continual learning. Fig.~\ref{fig3}a shows the MLP-Mixer architecture~\cite{tolstikhin2021mlp}, which consists of per-patch multilayer perceptrons (MLPs), mixer layers and an output layer, all implemented with fully connected layers (see Supplementary Fig.~2 for details). During initial learning, pretrained weights are programmed into the RM analogue array. The resulting conductance maps of the patch MLPs, mixer layer and output layer are shown in Fig.~\ref{fig3}a, with conductance values ranging from 10~$\mu$S to 80~$\mu$S. To mitigate RM programming overhead, subsequent machine unlearning of face ID~2 (Fig.~\ref{fig3}b) and continual learning of face ID~5 (Fig.~\ref{fig3}c) are performed by updating only the encoder and decoder parameters of the LoRA modules.

Fig.~\ref{fig3}d-f track the evolution of the representation space and recognition performance. Two-dimensional t-distributed stochastic neighbour embedding (t-SNE) visualizations of the embedding space reveal dynamic reconfiguration across the three stages. After initial learning, embeddings from five face classes form well-separated clusters, indicating accurate recognition (Fig.~\ref{fig3}d). After unlearning face ID~2 via gradient ascent, features of the second class (green) overlap with those of the fourth class (blue), indicating loss of discriminative power for the second identity (Fig.~\ref{fig3}e). Subsequent continual learning of a new face ID~5 using a replay strategy produces a new cluster that remains well separated from existing identities, demonstrating robust adaptability without catastrophic forgetting (Fig.~\ref{fig3}f).

Fig.~\ref{fig3}g highlights the benefits of the proposed hardware-software co-design by comparing inference accuracy obtained with fully analogue RM parameter updates and with the RM with digital LoRA accelerator (RM-DLoRA). The hybrid architecture achieves higher accuracy because it avoids repeated, error-prone analogue reprogramming, thereby reducing cumulative programming noise during inference after initial learning, machine unlearning (UL) and continual learning (CL) (see Supplementary Fig.~3 for the noise tolerance of RM programming).

A quantitative comparison of the three core workflows of training, deployment and inference is summarized in Fig.~\ref{fig3}h-j. Fig.~\ref{fig3}h shows that RM-DLoRA requires substantially lower training cost than conventional full-parameter fine-tuning on RM, reducing training update cost by 36.68$\times$ and 27.51$\times$ for UL and CL, respectively. Fig.~\ref{fig3}i quantifies write overhead during deployment. Whereas traditional RM program updates require repeated analogue reprogramming of RM arrays, RM-DLoRA updates only the LoRA weights stored in SRAM, reducing write-energy consumption by 92.91$\times$ and 69.68$\times$ for UL and CL, respectively. As shown in Fig.~\ref{fig3}j, RM-DLoRA further improves energy efficiency during inference procedure, achieving 21.77$\times$, 27.99$\times$ and 13.07$\times$ lower energy consumption than a GPU for learn, UL and CL, respectively (see Supplementary Table~1  for the energy breakdown).

\begin{figure}[!t]
\centering
\includegraphics[width=1.0\linewidth]{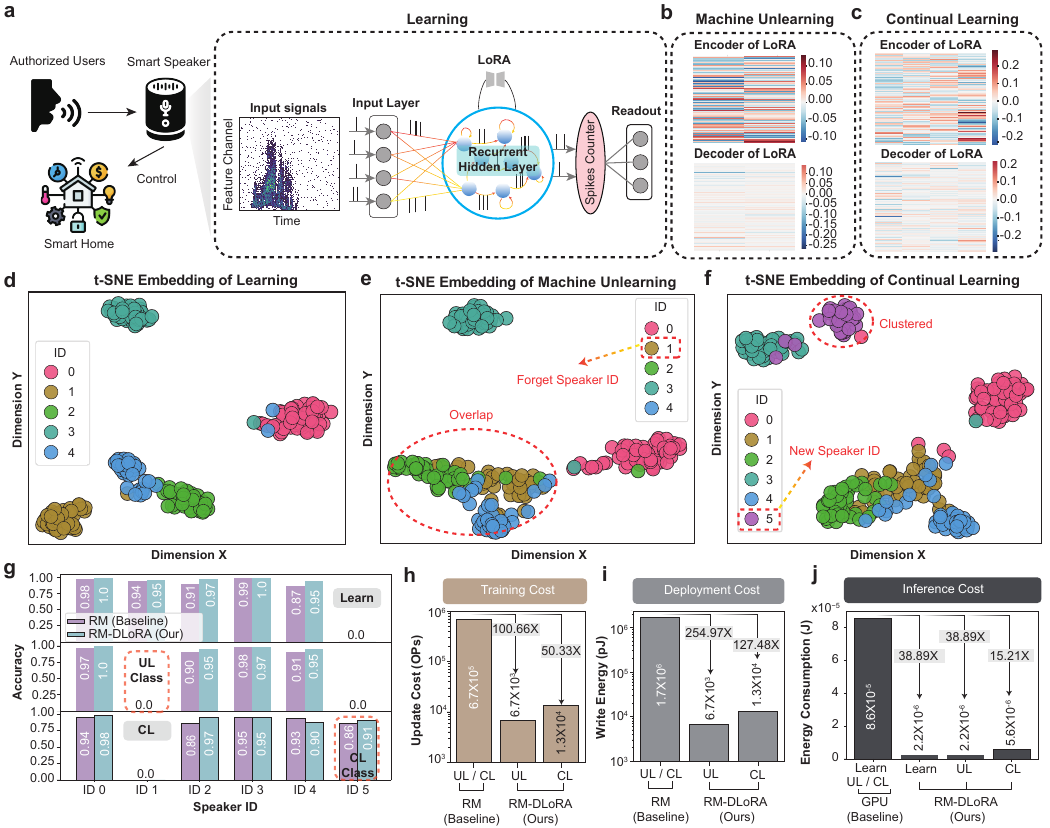}
\caption{\textbf{Learning, machine unlearning and continual learning in an RSNN for speaker-authentication system on the Spiking Speech Commands dataset.}
\textbf{a,} Schematic of the recurrent spiking neural network (RSNN) architecture. Audio signals from the Spiking Speech Commands dataset are converted into spike trains, accumulated over ten time windows and fed into the recurrent hidden layer of RSNN.
\textbf{b,} Distributions of LoRA encoder and decoder weights during machine unlearning of speaker ID~1.
\textbf{c,} Distributions of LoRA encoder and decoder weights during continual learning of a new speaker (ID~5).
\textbf{d-f,} Two-dimensional t-SNE visualizations of readout-layer activations after learning (\textbf{d}), machine unlearning (\textbf{e}) and continual learning (\textbf{f}).
\textbf{g,} Per-speaker classification accuracy for RM with digital LoRA updates (RM-DLoRA) and conventional RM program updates, showing that RM-DLoRA mitigates accuracy degradation caused by noise and conductance drift during repeated analogue reprogramming in learn, UL and CL tasks.
\textbf{h,} Training-update cost for RM-DLoRA and RM by full-parameter fine-tuning, with RM-DLoRA updates reducing the number of training update operations by about 100.66$\times$ and 50.33$\times$  for UL and CL tasks, respectively.
\textbf{i,} Write-energy consumption for RM-DLoRA and RM by full-parameter fine-tuning, with RM-DLoRA updates reducing write energy by about 254.97$\times$ and 127.48$\times$ for UL and CL tasks by avoiding repeated reprogramming of the RM array.
\textbf{j,} Inference energy for RM-DLoRA and a high-performance GPU, with RM-DLoRA achieving 38.89$\times$, 38.89$\times$ and 15.21$\times$ lower energy consumption across Learn, UL and CL.
\label{fig4}}
\end{figure}

\begin{figure}[!t]
\centering
\includegraphics[width=1.0\linewidth]{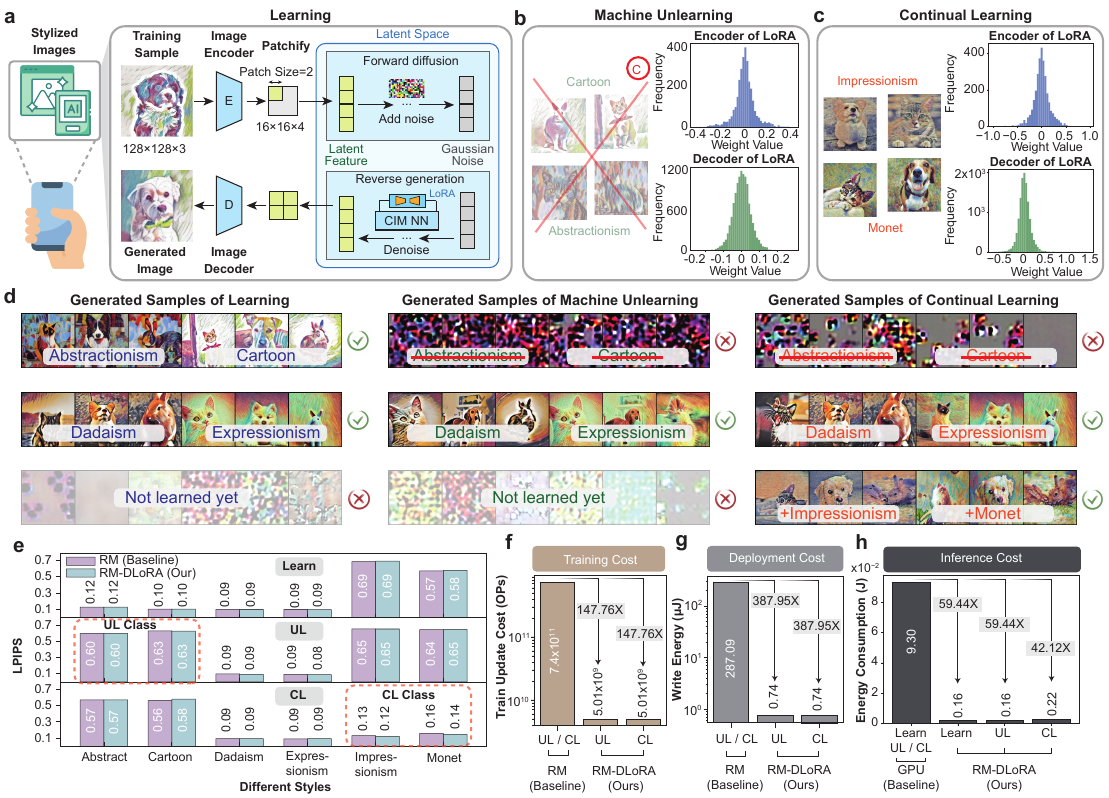}
\caption{\textbf{Learning, machine unlearning and continual learning in a conditional diffusion model on the UnlearnCanvas dataset.} 
\textbf{a,} Architecture of the latent diffusion model. RGB images of size 128$\times$128$\times$3 are encoded by a variational autoencoder (VAE) into 16$\times$16$\times$4 latent representations, which are partitioned into 2$\times$2 patches and fed as latent sequences to a DiT-B/2 backbone. Standard attention layers are replaced by depth-wise convolutional layers to improve efficiency. Image generation is performed by iterative denoising over 100 sampling steps. In the initial learning stage, the model is trained on four artistic styles (Expressionism, Dadaism, Cartoon and Abstractionism).
\textbf{b,} Machine unlearning of two styles (Cartoon and Abstractionism), revoking the corresponding generative capabilities.
\textbf{c,} Continual learning of two new styles (Impressionism and Monet), extending the model's repertoire without catastrophic forgetting.
\textbf{d,} Representative samples generated after learning, machine unlearning and continual learning, showing preservation of image content with appropriate application or suppression of the target styles.
\textbf{e,} Perceptual generation quality, measured by the Learned Perceptual Image Patch Similarity (LPIPS) metric (lower is better), comparing a full RM program-update baseline and RM-DLoRA; RM-DLoRA achieves comparable quality.
\textbf{f,} Training update cost across the UL and CL tasks, with RM-DLoRA reducing the number of trainable-parameter updates by about 147.76$\times$ relative to full-parameter fine-tuning for RM.
\textbf{g,} Write energy consumption, RM-DLoRA reducing write energy by about 387.95$\times$ for UL and CL by avoiding repeated analogue reprogramming of the RM array.
\textbf{h,} Inference energy consumption for RM-DLoRA and a high-performance GPU, showing 59.44$\times$, 59.44$\times$ and 42.12$\times$ lower energy for RM-DLoRA across learning (Learn), UL and CL tasks.
\label{fig5}}
\end{figure}

\subsection*{Speaker Authentication}

Speaker authentication is widely integrated into voice assistants on smart speakers to authenticate users and control smart home devices (Fig.~\ref{fig4}a). In such applications, enrolling new users and removing existing ones is routine. Therefore, the model should support efficient adaptation, including approximate unlearning of a removed user's voice and continual learning of newly enrolled speakers. Such capability is important for mitigating privacy-related risks, as it suppresses the model's ability to recognize speakers who should no longer be authorized. At the same time, the system must retain sufficient capacity to learn new voices without degrading recognition accuracy for previously registered users. We demonstrate these operations using the RM analogue--digital system with rank $r=8$ LoRA adapters.

Figs.~\ref{fig4}a-c illustrate the model architecture and parameter transformations during initial learning, machine unlearning and continual learning in a recurrent spiking neural network (RSNN) architecture inspired by liquid state machine~\cite{lin2025resistive} on speaker recognition system. Acoustic inputs are drawn from the Spiking Speech Commands dataset~\cite{cramer2020heidelberg}, which contains utterances from many speakers recorded under realistic, less-controlled conditions. As shown in Fig.~\ref{fig4}a, raw audio input signals are processed by the input layer and recurrent hidden layer of RSNN for temporal feature extraction, followed by a readout for classification. 
In the initial learning stage, the pretrained weights of the input, recurrent hidden and readout layers are programmed onto an RM analog array to support efficient inference. In the subsequent machine unlearning and continual learning processes, the LoRA weights are updated on a digital computer with an SRAM buffer as shown in Figs.~\ref{fig4}b and \ref{fig4}c .

Figs.~\ref{fig4}d-f present two-dimensional t-SNE visualizations of the readout layer activations after the three stages. After initial learning, activations from five enrolled speakers form distinct, well-separated clusters, indicating accurate speaker discrimination (Fig.~\ref{fig4}d). LoRA-based machine unlearning applied to speaker ID~1 causes its former cluster to overlap substantially with those of IDs~2 and 4 (Fig.~\ref{fig4}e), reflecting loss of discriminative power for the forgotten speaker while leaving the remaining speakers separable. Subsequent LoRA-based continual learning for a new speaker (ID~5) yields a compact, clearly separated cluster for the new identity while preserving the separation of the original speakers (Fig.~\ref{fig4}f).

Fig.~\ref{fig4}g compares inference accuracy across the three tasks (Learn, UL and CL). The RM-DLoRA architecture achieves higher per-speaker accuracy than conventional RM program updates method because it avoids repeated, error-prone analogue reprogramming, thereby reducing cumulative conductance drift in the RM array (see Supplementary Fig.~4 for the noise tolerance of RM programming).

Figs.~\ref{fig4}h-j quantify the benefits of the hardware-software co-design for training, on-device deployment and inference. Fig.~\ref{fig4}h shows that RM-DLoRA uses substantially fewer trainable parameters than the baseline full-parameter fine-tuning on RM array, reducing the number of training update operations by 100.66$\times$ and 50.33$\times$ across UL and CL tasks, respectively. Fig.~\ref{fig4}i compares write overhead: RM program updates require repeated reprogramming of the entire RM array, whereas our proposed RM-DLoRA updates only the LoRA weights stored in SRAM, reducing write-energy consumption by 254.97$\times$ and 127.48$\times$ for UL and CL tasks, respectively. As shown in Fig.~\ref{fig4}j, the hybrid architecture enables RM-DLoRA to achieve 38.89$\times$, 38.89$\times$ and 15.21$\times$ lower inference energy than a high-performance GPU for learning, machine unlearning and continual learning, respectively (see Supplementary Table~2  for the energy breakdown).

\subsection*{Stylized Image Generation}

Diffusion-based image generation has raised substantial societal concerns, including the creation of harmful content and copyright infringement~\cite{ringabell,zhang2024unlearncanvas}. In such settings, it is often necessary to suppress specific generative capabilities. However, access to the original training data may be restricted by data-governance, copyright, or deployment constraints, making full retraining from scratch impractical. Approximate machine unlearning therefore provides a practical mechanism for selectively reducing undesired behaviours. Similarly, when diffusion models must be adapted to generate images in new styles or under new conditions, continual learning mechanisms are required to expand model capabilities without catastrophic forgetting.

We demonstrate the effectiveness of our RM-based LoRA framework with LoRA rank $r=4$ for approximate machine unlearning, and continual learning in a conditional diffusion model. Performance is evaluated on the UnlearnCanvas dataset~\cite{zhang2024unlearncanvas}, which focuses on artistic style manipulation, with representative results shown in Fig.~\ref{fig5}.

Fig.~\ref{fig5}a-c illustrate the model architecture and task sequence. As shown in Fig.~\ref{fig5}a, the latent diffusion model takes 128$\times$128$\times$3 input images, which are compressed by a variational autoencoder into 16$\times$16$\times$4 latent representations. These latent maps are tokenized into 2$\times$2 patches and processed by a DiT-B/2 backbone~\cite{peebles2023scalable}, in which standard attention layers are replaced with depth-wise convolutions~\cite{liu2022convnet} to improve efficiency. The denoising process uses 100 sampling steps during inference to ensure high-quality generation (see Supplementary Fig.~5 for details of the model architecture). In the first phase (Fig.~\ref{fig5}b), the model learns four distinct artistic styles (Expressionism, Dadaism, Cartoon and Abstractionism). It then performs targeted machine unlearning of two styles (Cartoon and Abstractionism), mimicking scenarios in which particular generative capabilities must be revoked owing to licensing or ethical constraints. In the continual learning phase (Fig.~\ref{fig5}c), the model acquires two new styles (Impressionism and Monet), demonstrating the ability to accumulate new knowledge while retaining prior capabilities.

Representative samples across these stages (Fig.~\ref{fig5}d) show that the model preserves content fidelity while correctly applying or suppressing the specified styles. After unlearning, the removed styles no longer appear in generated outputs, providing strict control over prohibited categories (see Supplementary Fig.~6 for additional examples).

Quantitative results corroborate these observations. As shown in Fig.~\ref{fig5}e, the full RM program-update baseline achieves perceptual quality that is comparable to our RM-DLoRA method. Perceptual quality is quantified using the Learned Perceptual Image Patch Similarity (LPIPS) metric~\cite{zhang2018unreasonable} (lower is better). However, the performance of the full RM program-update baseline degrades severely in the presence of realistic programming noise, whereas RM-DLoRA maintains stable perceptual quality (see Supplementary Fig.~7 for the noise tolerance of RM programming). At the same time, the hybrid RM-DLoRA architecture delivers substantial efficiency gains over full-parameter fine-tuning. Fig.~\ref{fig5}f shows reductions in training update cost of 147.76$\times$ for UL and CL tasks, respectively. Fig.~\ref{fig5}g reports corresponding write energy savings of 387.95$\times$ for UL and CL tasks reflecting the avoidance of repeated analogue reprogramming of the RM array. Fig.~\ref{fig5}h shows that inference energy is reduced by 59.44$\times$, 59.44$\times$ and 42.12$\times$ relative to a high-performance GPU for learn, UL and CL tasks, respectively (see Supplementary Table~3  for the energy breakdown).

\section*{Discussion}

This study presents a hardware--software co-design framework that makes post-deployment model modification compatible with the physical constraints of RM-CIM hardware. Edge AI systems increasingly require local adaptation after deployment, including personalization, continual learning and approximate machine unlearning. However, these requirements conflict with the operating principles of RM-CIM accelerators, whose energy efficiency relies on keeping the analogue weight arrays largely fixed. Frequent reprogramming of RM devices introduces substantial write energy and latency overheads, is limited by device endurance, and may accumulate programming errors. Our RM-DLoRA framework addresses this mismatch by maintaining the pre-trained backbone weights in the analogue RM array while confining post-deployment updates to compact digital LoRA modules.

At the hardware level, this analogue--digital partition preserves the compute-in-memory advantage of RM arrays for high-throughput matrix--vector multiplication, while avoiding repeated analogue reprogramming during learning, approximate unlearning and continual learning. At the algorithmic level, low-rank adaptation restricts model updates to a small parameter subspace, substantially reducing the number of trainable parameters and update operations. As a result, RM-DLoRA enables efficient model modification without sacrificing the energy benefits of RM-CIM inference.

We validate this framework across both discriminative and generative workloads. For face and speaker recognition, RM-DLoRA supports the addition and removal of identities or speakers while preserving recognition performance for retained classes. Compared with full-parameter fine-tuning on RM hardware, it reduces training update count and write energy by orders of magnitude, and achieves substantially lower inference energy than high-performance GPUs, highlighting its suitability for privacy-sensitive edge applications. For conditional diffusion models evaluated on the UnlearnCanvas dataset, RM-DLoRA supports initial learning, approximate machine unlearning and continual learning of artistic styles while maintaining perceptual quality, as measured by LPIPS. These results indicate that the proposed framework is not limited to lightweight classifiers, but can also extend to more demanding generative models.

Importantly, the approximate unlearning demonstrated here should be interpreted as a practical mechanism for suppressing targeted behaviours, rather than as a formal privacy guarantee. In the generative setting, targeted approximate unlearning reduces the model's ability to produce selected styles, such as prohibited or undesired styles, thereby supporting licensing-, safety- and data-governance-related adaptation requirements. In recognition tasks, it reduces the influence of removed identities or speakers while maintaining performance on retained users. Future work may combine RM-DLoRA with certified unlearning, differential privacy or stronger auditing protocols to provide formal guarantees. More broadly, this work suggests that separating stable analogue backbone computation from lightweight digital adaptation is a promising direction for building energy-efficient, updateable and trustworthy edge intelligence.

\section*{Methods}

\subsection*{Fabrication of RM Chips}

The RM chips used in this work comprise a $128\times128$ 1T1R array fabricated in a \SI{180}{\nano\meter} CMOS technology node. Each 1T1R cell integrates an access transistor with an RM device formed between metal layers M5 and M6. The RM device stack adopts a TiN/TaO$_x$/TiN metal-insulator-metal (MIM) structure. The array is organized in a standard crossbar configuration with shared word lines (WLs), source lines (SLs), and bit lines (BLs). Specifically, the BLs are connected to the top electrodes of the RM devices, while the SLs are connected to the source terminals of the transistors. In this architecture, devices in the same row share a common BL and WL, whereas devices in the same column share a common SL.

\subsection*{Hybrid Analogue-digital Computing System}

The hybrid analogue-digital computing system is implemented on a custom printed circuit board (PCB) that integrates the \SI{180}{\nano\meter} RM CIM chip and peripheral circuits, and interfaces with a Xilinx ZC706 evaluation board based on a Zynq-7000 SoC. The digital control and data-processing functions are implemented in the programmable logic (FPGA fabric). The RM macro is addressed via serial-in/parallel-out shift registers (SN74HC595, Texas Instruments) and analogue multiplexers (CD4051B, Texas Instruments), while input voltages are provided by 16-bit digital-to-analogue converters (DAC80508, Texas Instruments). The output currents from the RM array are converted to voltages using transimpedance amplifiers (OPA4322, Texas Instruments) and subsequently digitized by 14-bit analogue-to-digital converters (ADS8324, Texas Instruments), with the digitized data processed by the FPGA fabric. A dedicated programming interface connects the RM array to a B1500A semiconductor device analyser for precise conductance initialization and reprogramming.

\subsection*{Machine Unlearning}

Machine unlearning aims to remove the influence of designated training examples from a pretrained model~\cite{bourtoule2021machine}. This is critical for mitigating unintended memorization and leakage of personally identifiable information~\cite{fredrikson2015model,carlini2019secret,carlini2021extracting}, as well as reducing the generation of unsafe outputs such as toxic or misleading content~\cite{wu2025unlearning,kumari2023ablating,zhang2024generate,zhang2024defensive}. By removing these contributions, unlearning supports compliance with the ``right to be forgotten'' under privacy-protecting regulations, including the EU General Data Protection Regulation~\cite{mantelero2013eu} and the California Consumer Privacy Act~\cite{ca_ab375_2018}.

Machine unlearning methods are commonly grouped into \emph{exact unlearning} and \emph{approximate unlearning}~\cite{jia2023model,liu2025rethinking,Philip2024}. Exact unlearning retrains the model from scratch after removing the target data, guaranteeing complete erasure but incurring prohibitive computational cost. Approximate unlearning uses more efficient techniques, such as gradient ascent~\cite{liu2022backdoor} or label obfuscation~\cite{Philip2024,graves2021amnesiac}, to approximate the effect of data removal without full retraining.

In this work we focus on efficient realization on RM-based accelerators rather than proposing new unlearning algorithms. We therefore adopt two representative approximate unlearning baselines:

\subsubsection*{Gradient Ascent Based Unlearning}

This approach removes the influence of the forget set \(\mathcal{D}_f\) by maximizing its loss, effectively reversing the original gradient descent updates. Starting from the pretrained parameters \(\theta_0\), the optimization objective is
\begin{equation}
\min_{\theta} \;
\underbrace{-\frac{1}{|\mathcal{D}_f|} \sum_{(x,y_f)\in \mathcal{D}_f} \ell(y_f \mid x; \theta)}_{\text{gradient ascent on forget set}}
\;+\;
\lambda \;
\underbrace{\frac{1}{|\mathcal{D}_r|} \sum_{(x,y)\in \mathcal{D}_r} \ell(y \mid x; \theta)}_{\text{gradient descent on retain set}},
\label{eq:mu-ga-unified}
\end{equation}
where \(\lambda \geq 0\) balances forgetting and retention, and \(\ell(\cdot;\theta)\) denotes the per-sample loss. The negative term induces gradient ascent on \(\mathcal{D}_f\), pushing predictions away from patterns induced by the forget data, while the retain term preserves performance on the retain set \(\mathcal{D}_r\).

\subsubsection*{Label Obfuscation Based Unlearning}

Another widely used technique replaces the original labels of samples in the forget set \(\mathcal{D}_f\) with random incorrect labels and fine-tunes the model on the combined dataset. The optimization objective is
\begin{equation}
\min_{\theta} \;
\underbrace{\frac{1}{|\mathcal{D}_f|} \sum_{(x,\tilde{y}_f)\in \tilde{\mathcal{D}}_f} \ell(\tilde{y}_f \mid x; \theta)}_{\text{gradient descent on forget set with label obfuscation}}
\;+\;
\lambda \;
\underbrace{\frac{1}{|\mathcal{D}_r|} \sum_{(x,y)\in \mathcal{D}_r} \ell(y \mid x; \theta)}_{\text{gradient descent on retain set}},
\label{eq:mu-rl-unified}
\end{equation}
where, for classification tasks,
\begin{equation}
\tilde{\mathcal{D}}_f
= \bigl\{(x, \tilde{y}_f) \,\big|\,
(x, y_f) \in \mathcal{D}_f,\;
\tilde{y}_f \sim \mathrm{Uniform}(\mathcal{Y} \setminus \{y_f\})\bigr\},
\end{equation}
is the forget set with randomized labels \(\tilde{y}_f\), and \(\lambda \geq 0\). This encourages the model to treat forgotten samples as belonging to arbitrary incorrect classes, thereby scrambling their original influence while preserving performance on the retain set \(\mathcal{D}_r\). For other types of tasks, the core idea of this method remains applicable, though its specific form may vary. For example, in generative tasks, when faced with conditions requiring unlearning, the network can be trained to shift its prediction target toward nonsensical or meaningless output. 

\subsection*{Continual Learning}

Continual learning methods can be broadly classified into three categories: \emph{regularization-based}~\cite{de2021continual,van2020brain}, \emph{replay-based}~\cite{rebuffi2017icarl,shin2017continual} and \emph{parameter-isolation-based} approaches~\cite{de2021continual}. These methods update model parameters to accommodate new tasks while preserving prior knowledge. In this work we only focus on efficient deployment on RM-based accelerators, and adopt a \emph{replay-based} continual learning framework as the baseline across all three application scenarios.

Replay-based methods maintain a compact memory buffer of past examples (or generate pseudo-samples) and jointly train them with current-task data to mitigate catastrophic forgetting. Let \(\theta_{t-1}\) denote the model parameters after learning the previous task and \(\ell(\cdot;\theta)\) the per-sample loss. Starting from \(\theta = \theta_{t-1}\), we optimize the following composite objective,
\begin{equation}
\min_{\theta} \;
\underbrace{\frac{1}{|\mathcal{D}_n|} \sum_{(x,y)\in \mathcal{D}_n} \ell(y \mid x; \theta)}_{\text{gradient descent on new data}}
\;+\;
\gamma \;
\underbrace{\frac{1}{|\mathcal{D}_p|} \sum_{(x,y)\in \mathcal{D}_p} \ell(y \mid x; \theta)}_{\text{gradient descent on replay buffer}},
\label{eq:cl-replay-unified}
\end{equation}
where \(\gamma \geq 0\) controls the trade-off between acquiring new knowledge and retaining prior knowledge, \(\mathcal{D}_n\) denotes the dataset of the current task and \(\mathcal{D}_p\) is the replay buffer storing representative samples from previous tasks (for example, selected via random sampling, herding or gradient-based prioritization). This joint optimization performs gradient descent on both sets, enabling the model to learn new task capabilities from \(\mathcal{D}_n\) while preserving performance on past tasks through \(\mathcal{D}_p\).

\subsection*{LoRA-based Machine Unlearning and Continual Learning}

To achieve parameter-efficient adaptation in both machine unlearning and continual learning, we incorporate LoRA~\cite{hu2022lora} into the gradient ascent based unlearning and replay-based continual learning frameworks introduced above. LoRA enables efficient fine-tuning by injecting trainable low-rank decomposition matrices into model layers while keeping the pretrained weights \(W_0 \in \mathbb{R}^{d \times k}\) frozen. The updated weight is expressed as
\begin{equation}
W = W_0 + \Delta W = W_0 + B A,
\label{eq:lora}
\end{equation}
where \(B \in \mathbb{R}^{d \times r}\), \(A \in \mathbb{R}^{r \times k}\), and \(r \ll \min(d, k)\) is the rank of the adaptation. Only the low-rank matrices \(A\) and \(B\) are optimized, reducing the number of trainable parameters from \(dk\) to \(r(d+k)\) per layer.

\subsubsection*{LoRA-based Machine Unlearning}

LoRA substantially reduces the computational and memory overhead of machine unlearning by updating only the LoRA matrices while keeping the pretrained weights \(W_0\) frozen. We apply LoRA to both gradient ascent based and label obfuscation based unlearning baselines, optimizing LoRA parameters \((A, B)\) (initialized as \(A_0, B_0\)) such that the forward pass uses the effective weights \(W_0 + BA\).

\paragraph{LoRA-based Gradient Ascent Unlearning.}

This variant adapts the gradient ascent objective from Equation~\eqref{eq:mu-ga-unified}. Optimization is performed solely over the adapter parameters,
\begin{equation}
\min_{A,B} \;
\underbrace{-\frac{1}{|\mathcal{D}_f|} \sum_{(x,y_f)\in \mathcal{D}_f} \ell(y_f \mid x; W_0 + BA)}_{\text{gradient ascent on forget set}}
\;+\;
\lambda \;
\underbrace{\frac{1}{|\mathcal{D}_r|} \sum_{(x,y)\in \mathcal{D}_r} \ell(y \mid x; W_0 + BA)}_{\text{gradient descent on retain set}}.
\label{eq:lora-ga-mu}
\end{equation}

The negative term induces gradient ascent on the forget set \(\mathcal{D}_f\) via the adapters, pushing the model away from patterns associated with the data to be forgotten, while the retain term preserves performance on \(\mathcal{D}_r\). 
We employ this approach for the face recognition task.

\paragraph{LoRA-based Label Obfuscation Unlearning.}

This method utilizes the label obfuscation Unlearning objective, Equation~\eqref{eq:mu-rl-unified} while training only the adapter parameters,
\begin{equation}
\min_{A,B} \;
\underbrace{\frac{1}{|\mathcal{D}_f|} \sum_{(x,\tilde{y}_f)\in \tilde{\mathcal{D}}_f} \ell(\tilde{y}_f \mid x; W_0 + BA)}_{\text{gradient descent on forget set with label obfuscation}}
\;+\;
\lambda \;
\underbrace{\frac{1}{|\mathcal{D}_r|} \sum_{(x,y)\in \mathcal{D}_r} \ell(y \mid x; W_0 + BA)}_{\text{gradient descent on retain set}},
\label{eq:lora-rl-mu}
\end{equation}
where, for classification tasks,
\begin{equation}
\tilde{\mathcal{D}}_f
= \bigl\{(x, \tilde{y}_f) \,\big|\,
(x, y_f) \in \mathcal{D}_f,\;
\tilde{y}_f \sim \mathrm{Uniform}(\mathcal{Y} \setminus \{y_f\})\bigr\},
\end{equation}
denotes the forget set with randomized incorrect labels \(\tilde{y}_f\). By fine-tuning only the adapters on this combined dataset, the model effectively erases the influence of the original forget samples while preserving its performance on the retain set. We apply this method to both speaker authentication and stylized image generation.


\subsubsection*{LoRA-based Continual Learning}

For replay-based continual learning, LoRA is similarly integrated into the joint training objective in Equation~\eqref{eq:cl-replay-unified}. At task \(t\), we initialize new LoRA adapters \((A_t, B_t)\) (or reuse and extend prior adapters) and optimize
\begin{equation}
\min_{A,B} \;
\frac{1}{|\mathcal{D}_n|} \sum_{(x,y)\in \mathcal{D}_n} \ell(y \mid x; W_0 + BA)
\;+\;
\gamma \;
\frac{1}{|\mathcal{D}_p|} \sum_{(x,y)\in \mathcal{D}_p} \ell(y \mid x; W_0 + BA).
\label{eq:lora-cl}
\end{equation}

Both terms involve \emph{gradient descent}: the first adapts the model to the new task \(\mathcal{D}_n\), and the second reinforces prior knowledge via the replay buffer \(\mathcal{D}_p\). By updating only the LoRA adapters, this technique enables efficient continual learning across face recognition, speaker authentication and stylized image generation tasks without modifying the pretrained weights, mitigating catastrophic forgetting while minimizing resource demands.


\section*{Data Availability}

The Olivetti faces dataset~\cite{samaria1994parameterisation}, the UnlearnCanvas dataset~\cite{zhang2024unlearncanvas} and Spiking Speech Commands dataset~\cite{cramer2020heidelberg} are publicly available.

\section*{Code Availability}
The code that supports the plots within this paper is available at \href{https://github.com/MrLinNing/RMAdaptiveMachine}{https://github.com/MrLinNing/RMAdaptiveMachine}. 

\section*{Acknowledgement}
This research is supported by the National Natural Science Foundation of China (Grant Nos. 62374181, 62488101, 62495104), Hong Kong Research Grant Council (Grant No. 17212923, C1009-22G, C7003-24Y and AOE/E-101/23-N). This research is also partially supported by ACCESS - AI Chip Center for Emerging Smart Systems, sponsored by Innovation and Technology Fund (ITF), Hong Kong SAR.

\section*{Author Contributions}

N.L., JC.Y., YG.H. contributed to the design and development of the models, software, and hardware experiments. All authors discussed the results and implications and commented on the manuscript at all stages.

\section*{Competing Interests}
The authors declare no competing interests.


\bibliography{reference}

@article{wu2026hardware,
  title={Hardware-aware Low-Rank Adaptation for Large Language Models Based on Hybrid Compute-in-Memory Architecture},
  author={Wu, Taiqiang and Ding, Chenchen and Zhou, Wenyong and Cheng, Yuxin and Feng, Xincheng and Wang, Shuqi and Xu, Wendong and Shi, Chufan and Liu, Zhengwu and Wong, Ngai},
  journal={ACM Transactions on Design Automation of Electronic Systems},
  volume={31},
  number={5},
  pages={1--23},
  year={2026},
  publisher={ACM New York, NY}
}

@article{wang2024safe,
  title={Safe, secure and trustworthy compute-in-memory accelerators},
  author={Wang, Ziyu and Wu, Yuting and Park, Yongmo and Lu, Wei D},
  journal={Nature Electronics},
  volume={7},
  number={12},
  pages={1086--1097},
  year={2024},
  publisher={Nature Publishing Group UK London}
}

@inproceedings{lin2025guarder,
  title={Guarder: A Stable and Lightweight Reconfigurable RRAM-based PIM Accelerator for DNN IP Protection},
  author={Lin, Ning and Li, Yi and Li, Jiankun and Yang, Jichang and He, Yangu and Luo, Yukui and Shang, Dashan and Chen, Xiaoming and Qi, Xiaojuan and Wang, Zhongrui},
  booktitle={2025 62nd ACM/IEEE Design Automation Conference (DAC)},
  pages={1--7},
  year={2025},
  organization={IEEE}
}

@article{liu2026privacy,
  title={Privacy-preserving data analysis using a memristor chip with colocated authentication and processing},
  author={Liu, Zhengwu and Wang, Zhongrui and Ding, Chenchen and Lin, Bohan and Tang, Jianshi and Gao, Bin and Wong, Ngai and Wu, Huaqiang},
  journal={Science Advances},
  volume={12},
  number={6},
  pages={eady5485},
  year={2026},
  publisher={American Association for the Advancement of Science}
}

@article{yue2025physical,
  title={Physical unclonable in-memory computing for simultaneous protecting private data and deep learning models},
  author={Yue, Wenshuo and Wu, Kai and Li, Zhiyuan and Zhou, Juchen and Wang, Zeyu and Zhang, Teng and Yang, Yuxiang and Ye, Lintao and Wu, Yongqin and Bu, Weihai and others},
  journal={Nature Communications},
  volume={16},
  number={1},
  pages={1031},
  year={2025},
  publisher={Nature Publishing Group UK London}
}

@article{ahmed2026mof,
  title={MoF-LoRA: Mixture of Low-Rank Fault-Tolerant Experts for RRAM-based In-Memory Computing},
  author={Ahmed, Soyed Tuhin and Ortega, Eduardo and Xiao, Patrick and Feinberg, Ben and Bennett, Christopher H and Marinella, Matthew J and Chakrabarty, Krishnendu},
  journal={IEEE Journal on Emerging and Selected Topics in Circuits and Systems},
  year={2026},
  publisher={IEEE}
}

@article{cramer2020heidelberg,
  title={The heidelberg spiking data sets for the systematic evaluation of spiking neural networks},
  author={Cramer, Benjamin and Stradmann, Yannik and Schemmel, Johannes and Zenke, Friedemann},
  journal={IEEE Transactions on Neural Networks and Learning Systems},
  volume={33},
  number={7},
  pages={2744--2757},
  year={2020},
  publisher={IEEE}
}

@article{wistuba2023continual,
  title={Continual learning with low rank adaptation},
  author={Wistuba, Martin and Sivaprasad, Prabhu Teja and Balles, Lukas and Zappella, Giovanni},
  journal={arXiv preprint arXiv:2311.17601},
  year={2023}
}

@inproceedings{bourtoule2021machine,
  title={Machine unlearning},
  author={Bourtoule, Lucas and Chandrasekaran, Varun and Choquette-Choo, Christopher A and Jia, Hengrui and Travers, Adelin and Zhang, Baiwu and Lie, David and Papernot, Nicolas},
  booktitle={2021 IEEE symposium on security and privacy (SP)},
  pages={141--159},
  year={2021},
  organization={IEEE}
}

@article{mantelero2013eu,
  title={The eu proposal for a general data protection regulation and the roots of the ‘right to be forgotten’},
  author={Mantelero, Alessandro},
  journal={Computer Law \& Security Review},
  volume={29},
  number={3},
  pages={229--235},
  year={2013},
  publisher={Elsevier}
}

@misc{ca_ab375_2018,
  title    = {Assembly Bill No. 375: An act to add Title 1.81.5 (commencing with Section 1798.100) to Part 4 of Division 3 of the Civil Code, relating to privacy},
  author   = {Chau},
  year     = {2018},
  url      = {https://leginfo.legislature.ca.gov/faces/billTextClient.xhtml?bill_id=201720180AB375},
  note     = {California Consumer Privacy Act of 2018 (Chapter 55, Statutes of 2018)},
  urldate  = {2025-10-15}
}

@article{liu2025rethinking,
  title={Rethinking machine unlearning for large language models},
  author={Liu, Sijia and Yao, Yuanshun and Jia, Jinghan and Casper, Stephen and Baracaldo, Nathalie and Hase, Peter and Yao, Yuguang and Liu, Chris Yuhao and Xu, Xiaojun and Li, Hang and others},
  journal={Nature Machine Intelligence},
  pages={1--14},
  year={2025},
  publisher={Nature Publishing Group UK London}
}

@article{jia2023model,
  title={Model sparsity can simplify machine unlearning},
  author={Jia, Jinghan and Liu, Jiancheng and Ram, Parikshit and Yao, Yuguang and Liu, Gaowen and Liu, Yang and Sharma, Pranay and Liu, Sijia},
  journal={Advances in Neural Information Processing Systems},
  volume={36},
  pages={51584--51605},
  year={2023}
}

@inproceedings{liu2022backdoor,
  title={Backdoor defense with machine unlearning},
  author={Liu, Yang and Fan, Mingyuan and Chen, Cen and Liu, Ximeng and Ma, Zhuo and Wang, Li and Ma, Jianfeng},
  booktitle={IEEE INFOCOM 2022-IEEE conference on computer communications},
  pages={280--289},
  year={2022},
  organization={IEEE}
}

@inproceedings{graves2021amnesiac,
  title={Amnesiac machine learning},
  author={Graves, Laura and Nagisetty, Vineel and Ganesh, Vijay},
  booktitle={Proceedings of the AAAI Conference on Artificial Intelligence},
  pages={11516--11524},
  year={2021}
}

@article{Philip2024,
author = {Xu, Heng and Zhu, Tianqing and Zhang, Lefeng and Zhou, Wanlei and Yu, Philip S.},
title = {Machine Unlearning: A Survey},
year = {2023},
issue_date = {January 2024},
publisher = {Association for Computing Machinery},
address = {New York, NY, USA},
volume = {56},
number = {1},
issn = {0360-0300},
url = {https://doi.org/10.1145/3603620},
doi = {10.1145/3603620},
journal = {ACM Comput. Surv.},
month = aug,
articleno = {9},
numpages = {36}
}

@article{lin2025resistive,
  title={Resistive memory-based zero-shot liquid state machine for multimodal event data learning},
  author={Lin, Ning and Wang, Shaocong and Li, Yi and Wang, Bo and Shi, Shuhui and He, Yangu and Zhang, Woyu and Yu, Yifei and Zhang, Yue and Zhang, Xinyuan and others},
  journal={Nature Computational Science},
  volume={5},
  number={1},
  pages={37--47},
  year={2025},
  publisher={Nature Publishing Group US New York}
}

@inproceedings{hu2022lora,
  author    = {Hu, Edward J. and Shen, Yelong and Wallis, Phillip and Allen-Zhu, Zeyuan and Li, Yuanzhi and Wang, Shean and Wang, Lu and Chen, Weizhu},
  title     = {LoRA: Low-Rank Adaptation of Large Language Models},
  booktitle = {International Conference on Learning Representations},
  year      = {2022},
  url       = {https://openreview.net/forum?id=nZeVKeeFYf9}
}

@article{de2021continual,
  title={A continual learning survey: Defying forgetting in classification tasks},
  author={De Lange, Matthias and Aljundi, Rahaf and Masana, Marc and Parisot, Sarah and Jia, Xu and Leonardis, Ale{v{s}} and Slabaugh, Gregory and Tuytelaars, Tinne},
  journal={IEEE transactions on pattern analysis and machine intelligence},
  volume={44},
  number={7},
  pages={3366--3385},
  year={2021},
  publisher={IEEE}
}

@article{shin2017continual,
  title={Continual learning with deep generative replay},
  author={Shin, Hanul and Lee, Jung Kwon and Kim, Jaehong and Kim, Jiwon},
  journal={Advances in neural information processing systems},
  volume={30},
  year={2017}
}

@inproceedings{rebuffi2017icarl,
  title={icarl: Incremental classifier and representation learning},
  author={Rebuffi, Sylvestre-Alvise and Kolesnikov, Alexander and Sperl, Georg and Lampert, Christoph H},
  booktitle={Proceedings of the IEEE conference on Computer Vision and Pattern Recognition},
  pages={2001--2010},
  year={2017}
}

@article{van2020brain,
  title={Brain-inspired replay for continual learning with artificial neural networks},
  author={Van de Ven, Gido M and Siegelmann, Hava T and Tolias, Andreas S},
  journal={Nature communications},
  volume={11},
  number={1},
  pages={4069},
  year={2020},
  publisher={Nature Publishing Group UK London}
}

@article{lanza2025growing,
  title={The growing memristor industry},
  author={Lanza, Mario and Pazos, Sebastian and Aguirre, Fernando and Sebastian, Abu and Le Gallo, Manuel and Alam, Syed M and Ikegawa, Sumio and Yang, J Joshua and Vianello, Elisa and Chang, Meng-Fan and others},
  journal={Nature},
  volume={640},
  number={8059},
  pages={613--622},
  year={2025},
  publisher={Nature Publishing Group UK London}
}

@article{wan2022compute,
  title={A compute-in-memory chip based on resistive random-access memory},
  author={Wan, Weier and Kubendran, Rajkumar and Schaefer, Clemens and Eryilmaz, Sukru Burc and Zhang, Wenqiang and Wu, Dabin and Deiss, Stephen and Raina, Priyanka and Qian, He and Gao, Bin and others},
  journal={Nature},
  volume={608},
  number={7923},
  pages={504--512},
  year={2022},
  publisher={Nature Publishing Group UK London}
}

@article{yao2020fully,
  title={Fully hardware-implemented memristor convolutional neural network},
  author={Yao, Peng and Wu, Huaqiang and Gao, Bin and Tang, Jianshi and Zhang, Qingtian and Zhang, Wenqiang and Yang, J Joshua and Qian, He},
  journal={Nature},
  volume={577},
  number={7792},
  pages={641--646},
  year={2020},
  publisher={Nature Publishing Group UK London}
}

@article{liu2022optoelectronic,
  title={An optoelectronic synapse based on $\alpha$-In2Se3 with controllable temporal dynamics for multimode and multiscale reservoir computing},
  author={Liu, Keqin and Zhang, Teng and Dang, Bingjie and Bao, Lin and Xu, Liying and Cheng, Caidie and Yang, Zhen and Huang, Ru and Yang, Yuchao},
  journal={Nature Electronics},
  volume={5},
  number={11},
  pages={761--773},
  year={2022},
  publisher={Nature Publishing Group UK London}
}

@article{yangdong2025ultrahigh,
  title={Ultrahigh-precision analog computing using memory-switching geometric ratio of transistors},
  author={Yangdong, Xing-Jian and Wang, Cong and Zhao, Yichen and Wang, Zi-Chun and Yang, Zaizheng and Liu, Zenglin and Yu, Wentao and Zeng, Zhoujie and Wang, Shuang and Wei, Wei and others},
  journal={Science Advances},
  volume={11},
  number={37},
  pages={eady4798},
  year={2025},
  publisher={American Association for the Advancement of Science}
}

@article{rao2023thousands,
  title={Thousands of conductance levels in memristors integrated on CMOS},
  author={Rao, Mingyi and Tang, Hao and Wu, Jiangbin and Song, Wenhao and Zhang, Max and Yin, Wenbo and Zhuo, Ye and Kiani, Fatemeh and Chen, Benjamin and Jiang, Xiangqi and others},
  journal={Nature},
  volume={615},
  number={7954},
  pages={823--829},
  year={2023},
  publisher={Nature Publishing Group UK London}
}

@article{song2024programming,
  title={Programming memristor arrays with arbitrarily high precision for analog computing},
  author={Song, Wenhao and Rao, Mingyi and Li, Yunning and Li, Can and Zhuo, Ye and Cai, Fuxi and Wu, Mingche and Yin, Wenbo and Li, Zongze and Wei, Qiang and others},
  journal={Science},
  volume={383},
  number={6685},
  pages={903--910},
  year={2024},
  publisher={American Association for the Advancement of Science}
}

@article{zuo2025precise,
  title={Precise and scalable analogue matrix equation solving using resistive random-access memory chips},
  author={Zuo, Pushen and Wang, Qishen and Luo, Yubiao and Xie, Ruiqing and Wang, Shiqing and Cheng, Zezhi and Bao, Lin and Wang, Zongwei and Cai, Yimao and Huang, Ru and others},
  journal={Nature Electronics},
  pages={1--12},
  year={2025},
  publisher={Nature Publishing Group UK London}
}

@article{ringabell,
  title={Ring-a-bell! how reliable are concept removal methods for diffusion models?},
  author={Tsai, Yu-Lin and Hsu, Chia-Yi and Xie, Chulin and Lin, Chih-Hsun and Chen, Jia-You and Li, Bo and Chen, Pin-Yu and Yu, Chia-Mu and Huang, Chun-Ying},
  journal={arXiv preprint arXiv:2310.10012},
  year={2023}
}

@inproceedings{fredrikson2015model,
  title={Model inversion attacks that exploit confidence information and basic countermeasures},
  author={Fredrikson, Matt and Jha, Somesh and Ristenpart, Thomas},
  booktitle={Proceedings of the 22nd ACM SIGSAC conference on computer and communications security},
  pages={1322--1333},
  year={2015}
}

@inproceedings{carlini2019secret,
  title={The secret sharer: Evaluating and testing unintended memorization in neural networks},
  author={Carlini, Nicholas and Liu, Chang and Erlingsson, {\'U}lfar and Kos, Jernej and Song, Dawn},
  booktitle={28th USENIX security symposium (USENIX security 19)},
  pages={267--284},
  year={2019}
}

@inproceedings{carlini2021extracting,
  title={Extracting training data from large language models},
  author={Carlini, Nicholas and Tramer, Florian and Wallace, Eric and Jagielski, Matthew and Herbert-Voss, Ariel and Lee, Katherine and Roberts, Adam and Brown, Tom and Song, Dawn and Erlingsson, Ulfar and others},
  booktitle={30th USENIX security symposium (USENIX Security 21)},
  pages={2633--2650},
  year={2021}
}

@inproceedings{wu2025unlearning,
  title={Unlearning concepts in diffusion model via concept domain correction and concept preserving gradient},
  author={Wu, Yongliang and Zhou, Shiji and Yang, Mingzhuo and Wang, Lianzhe and Chang, Heng and Zhu, Wenbo and Hu, Xinting and Zhou, Xiao and Yang, Xu},
  booktitle={Proceedings of the AAAI Conference on Artificial Intelligence},
  pages={8496--8504},
  year={2025}
}

@inproceedings{kumari2023ablating,
  title={Ablating concepts in text-to-image diffusion models},
  author={Kumari, Nupur and Zhang, Bingliang and Wang, Sheng-Yu and Shechtman, Eli and Zhang, Richard and Zhu, Jun-Yan},
  booktitle={Proceedings of the IEEE/CVF International Conference on Computer Vision},
  pages={22691--22702},
  year={2023}
}

@inproceedings{zhang2024generate,
  title={To generate or not? safety-driven unlearned diffusion models are still easy to generate unsafe images... for now},
  author={Zhang, Yimeng and Jia, Jinghan and Chen, Xin and Chen, Aochuan and Zhang, Yihua and Liu, Jiancheng and Ding, Ke and Liu, Sijia},
  booktitle={European Conference on Computer Vision},
  pages={385--403},
  year={2024},
  organization={Springer}
}

@article{zhang2024defensive,
  title={Defensive unlearning with adversarial training for robust concept erasure in diffusion models},
  author={Zhang, Yimeng and Chen, Xin and Jia, Jinghan and Zhang, Yihua and Fan, Chongyu and Liu, Jiancheng and Hong, Mingyi and Ding, Ke and Liu, Sijia},
  journal={Advances in neural information processing systems},
  volume={37},
  pages={36748--36776},
  year={2024}
}

@inproceedings{samaria1994parameterisation,
  title={Parameterisation of a stochastic model for human face identification},
  author={Samaria, Ferdinando S and Harter, Andy C},
  booktitle={Proceedings of 1994 IEEE workshop on applications of computer vision},
  pages={138--142},
  year={1994},
  organization={IEEE}
}

@article{tolstikhin2021mlp,
  title={Mlp-mixer: An all-mlp architecture for vision},
  author={Tolstikhin, Ilya O and Houlsby, Neil and Kolesnikov, Alexander and Beyer, Lucas and Zhai, Xiaohua and Unterthiner, Thomas and Yung, Jessica and Steiner, Andreas and Keysers, Daniel and Uszkoreit, Jakob and others},
  journal={Advances in neural information processing systems},
  volume={34},
  pages={24261--24272},
  year={2021}
}

@inproceedings{peebles2023scalable,
  title={Scalable diffusion models with transformers},
  author={Peebles, William and Xie, Saining},
  booktitle={Proceedings of the IEEE/CVF international conference on computer vision},
  pages={4195--4205},
  year={2023}
}

@inproceedings{liu2022convnet,
  title={A convnet for the 2020s},
  author={Liu, Zhuang and Mao, Hanzi and Wu, Chao-Yuan and Feichtenhofer, Christoph and Darrell, Trevor and Xie, Saining},
  booktitle={Proceedings of the IEEE/CVF conference on computer vision and pattern recognition},
  pages={11976--11986},
  year={2022}
}

@inproceedings{zhang2018unreasonable,
  title={The unreasonable effectiveness of deep features as a perceptual metric},
  author={Zhang, Richard and Isola, Phillip and Efros, Alexei A and Shechtman, Eli and Wang, Oliver},
  booktitle={Proceedings of the IEEE conference on computer vision and pattern recognition},
  pages={586--595},
  year={2018}
}

@article{zhang2024unlearncanvas,
  title={Unlearncanvas: A stylized image dataset to benchmark machine unlearning for diffusion models},
  author={Zhang, Yihua and Zhang, Yimeng and Yao, Yuguang and Jia, Jinghan and Liu, Jiancheng and Liu, Xiaoming and Liu, Sijia},
  journal={CoRR},
  year={2024}
}

\end{document}